\documentclass[aps,pre,twocolumn,groupedaddress,amsmath,amssymb]{revtex4-1} 
\usepackage{amssymb,amsfonts,amsmath}
\usepackage{graphicx}
\usepackage{amsbsy}
\usepackage{revsymb}
\usepackage{color}
\usepackage{dcolumn}
\usepackage{bm}

\def\C{{\@QC C}}
\def\@QC#1{\mathpalette{\setbox0=\hbox\bgroup$\rm}%
  {\egroup C$\egroup\rm\rlap{\kern0.4\wd0\vrule
  width 0.05\wd0 height 0.97\ht0 depth -0.01\ht0}%
  #1\bgroup}}

\usepackage[latin1]{inputenc} 

\begin{document}

\title{Quantum Ring-Polymer Contraction Method: Including nuclear quantum effects at no additional computational cost in comparison to \textit{ab-initio} molecular dynamics}

\author{Christopher John}
\affiliation{Dynamics of Condensed Matter, Department of Chemistry, University of Paderborn, Warburger Str. 100, D-33098 Paderborn, Germany}
\author{Thomas Spura}
\affiliation{Dynamics of Condensed Matter, Department of Chemistry, University of Paderborn, Warburger Str. 100, D-33098 Paderborn, Germany}
\author{Scott Habershon}
\affiliation{Department of Chemistry and Centre for Scientific Computing, University of Warwick, Coventry CV4 7AL, United Kingdom.}
\author{Thomas D. K\"uhne}
\email{tdkuehne@mail.upb.de}
\affiliation{Dynamics of Condensed Matter, Department of Chemistry, University of Paderborn, Warburger Str. 100, D-33098 Paderborn, Germany}
\affiliation{Paderborn Center for Parallel Computing and Institute for Lightweight Design, Department of Chemistry, University of Paderborn, Warburger Str. 100, D-33098 Paderborn, Germany}

\date{\today}

\begin{abstract}
We present a simple and accurate computational method, which facilitates \textit{ab-initio} path-integral molecular dynamics simulations, where the quantum mechanical nature of the nuclei is explicitly taken into account, at essentially no additional computational cost in comparison to the corresponding calculation using classical nuclei. The predictive power of the proposed quantum ring-polymer contraction method is demonstrated by computing various static and dynamic properties of liquid water at ambient conditions. This development permits to routinely include nuclear quantum effects in \textit{ab-initio} molecular dynamics simulations.
\end{abstract}

\maketitle











\section{Introduction}

A large variety of relevant problems in chemical physics are dominated by nuclear quantum effects (NQE), such as zero-point energy (ZPE) and tunnelling \cite{BernePIreview}. In particular, at low temperature and for systems containing light atoms, NQE must be considered to obtain the correct quantitative, and sometimes even qualitative behavior. Path-integral molecular dynamics (PIMD) simulations are a simple and computationally efficient option to calculating quantum-mechanical time-independent properties in many-particle systems \cite{Feynman, chandler1981isomorphism, parrinello1984moltenkcl}. As such, PIMD methods are the standard approach for understanding the role of NQE in complex condensed-phase systems, with examples ranging from the analysis of quantum fluctuations in water and ice 
\cite{kuharski1085studystructure, wallqvist1985pimd, lobaugh1997quantumwater, Mahoney2001, hernandez2006modeldependence, Morrone2007, comp_quant_eff, Fanourgakis2009, Vega2010, PhysRevLett.107.145501, HabershonManolopoulos2011, PhysRevLett.109.226101, MarklandPNAS2012, CeriottiJPCC2013, NqePNAS2013, kuehneJACS2014, Fritsch2014, Medders2014, SpuraCC2015, DelleSite2015, Kessler2015} 
to simulations of proton transfer in biological enzyme catalysis \cite{Warshel1993, Gao2002, Warshel2004, Gao2008, MillerPNAS2011}, just to name a few. However, in the past decade or so, PIMD-based approaches have been developed, which also allow the approximation of time-dependent properties, such as diffusion coefficients and vibrational spectra; methods such as centroid molecular dynamics (CMD) \cite{CMD1993, hone:154103} and ring-polymer molecular dynamics
(RPMD) \cite{craig:3368, RPMDreview}, therefore, provide a tractable route to quantum dynamics in liquids and solids \cite{miller2005Hquantumdiffusion, markland:194506, habershonIR, Bowman2008, habershon2009quantumleakage, WhMiller2009, Paesani2010, Rossi2014, SpuraWater2015, Medders2015}. 

In general, path-integral simulations can be performed using an arbitrary interaction potential $V(\mathbf{R})$, where $\mathbf{R}=\{ \mathbf{R}_I \}$ are the nuclear coordinates; for instance, simulations of liquid water have employed both empirical force-fields \cite{kuharski1085studystructure, wallqvist1985pimd, lobaugh1997quantumwater, Mahoney2001, hernandez2006modeldependence, Fanourgakis2006, comp_quant_eff, Vega2010, Medders2014, SpuraWater2015} and density functional theory (DFT) \cite{PhysRevLett.91.215503, PhysRevLett.101.017801,LinAI2014} to describe the interactions between the atoms. However, because each quantum particle is replaced by a classical $P$-bead harmonic ring-polymer in the path-integral approach, the computational expense required to evaluate $V(\mathbf{R})$ and the nuclear forces $\mathbf{F}_I=-\nabla_{\mathbf{R}_I} V(\mathbf{R})$ in the extended path-integral phase-space is roughly $P$ times greater than the analogue classical molecular dynamics (MD) simulation. Thus, even at room temperature, PIMD simulations tend to be computationally at least one or two orders of magnitude more expensive than classical MD; in the case of liquid water under ambient conditions, it is found that $P=32$ ring-polymer beads are required to converge most calculated properties of interest \cite{comp_quant_eff, SpuraWater2015}. As a result of the increased computational cost of PIMD simulations, the vast majority of such calculations employ empirical force-fields to model the intra- and intermolecular interactions. By contrast, 
using an accurate electronic structure theory-based \textit{ab-initio} potential to compute the nuclear forces, the resulting \textit{ab-initio} PIMD simulations are severely limited with regard to the attainable length and time scales \cite{marx:4077, tuckerman:5579, MarxJPCA2009, SpuraCC2015}. 
Therefore, quite a few methods have been recently developed to accelerate the convergence with respect to the number of ring-polymer beads. Particularly notable examples for such acceleration techniques are higher-order Trotter factorization approaches \cite{TakahashiImada1984, LiBroughton1987, Suzuki1990, Suzuki1992, Chin1997, JangVoth2001, PerezJCP2011, Chin2015}, as well as schemes based on the generalized Langevin equation \cite{ceriotti2009nuclearquantumeffects, CeriottiJCTC2010, CeriottiJCP2011, PhysRevLett.109.100604}. Unfortunately, both of these acceleration methods are limited to the calculation of quantum mechanical, though static equilibrium properties only. Nevertheless, it is possible to bypass this shortcoming using the recently devised ring-polymer contraction (RPC) scheme \cite{markland:024105, Markland2008256}. However, due to the fact the latter approach relies on the notion that the interaction potential can be conveniently decomposed into a rapidly varying hard and a more slowly varying soft part, it can so far only be employed in simulations with empirical force-fields, since no such decomposition of an \textit{ab-initio} potential is not known as yet \cite{RPMDreview}. 

In this work, we propose a simple and accurate quantum RPC (q-RPC) method to facilitate \textit{ab-initio} PIMD simulations at essentially no additional computational cost in comparison to conventional \textit{ab-initio} MD \cite{Marx2009, CP2Greview}. To that extent, inspired by the original RPC scheme \cite{markland:024105, Markland2008256}, the \textit{ab-initio} potential $V_{AI}(\mathbf{R})$ is decomposed into a slowly varying soft and a much more quickly varying hard potential by means of a computationally inexpensive auxiliary potential $V_{aux}(\mathbf{R})$, i.e. $V_{AI}(\mathbf{R}) = V_{AI}(\mathbf{R}) - V_{aux}(\mathbf{R}) + V_{aux}(\mathbf{R})$. Specifically, $V_{aux}(\mathbf{R})$ can be though of as the hard part of the interaction potential, whereas the difference potential $V_{AI}(\mathbf{R}) - V_{aux}(\mathbf{R})$ can be considered as soft. 
As a consequence, the former must be evaluated using the full number of beads $P$, while the computationally expensive difference potential can be computed on a much smaller ring-polymer with just $P^{\prime} < P$ beads. Given that $V_{aux}(\mathbf{R})$ roughly resembles $V_{AI}(\mathbf{R})$, we found that $P^{\prime} = 1$ is typically sufficient to yield accurate results. 

The remainder of the paper is organized as follows. In section II, we outline the relevant theory relating to path-integral and RPC simulation methods. In section III, the present q-RPC method is described in detail. Thereafter, in section IV, we assess the efficiency and accuracy of the q-RPC approach via \textit{ab-initio} PIMD simulations of static and dynamic properties of bulk liquid water at ambient conditions, before concluding the paper in section V.

\section{Path-Integral Formalism}

Before introducing the q-RPC method in the next section, we briefly summarize the key principles of the path-integral formalism that are at the bases of our novel approach.

\subsection{Path-Integral Molecular Dynamics}

In the PIMD formalism, every quantum particle is substituted by a harmonic $P$-bead ring-polymer that is subsequently treated classically. In other words, since the extended ring-polymer system is isomorphic to the original quantum system, it possible to compute quantum mechanical properties essentially exactly by sampling the extended path-integral phase space using MD \cite{Feynman, chandler1981isomorphism, parrinello1984moltenkcl}. On that point, the canonical quantum partition function $Z(\beta)$ is written in terms of Hamilton operator $\hat{H}=\hat{T}+\hat{V}$, i.e. 
\begin{equation}
  \label{PartitionFunction}
  Z(\beta)=\textrm{Tr} \left [ e^{-\beta\hat{H}}\right ] = \textrm{Tr}
  \left [ \left(e^{-\beta_{P}\hat{H}} \right)^{P}  \right ] =\lim_{{P \to \infty}}
  Z_{P}(\beta),
\end{equation}
where $\beta^{-1}=k_{B}T$ is the inverse temperature. This implies that the origin of path-integral formalism is the notion that the finite temperature density matrix $e^{-\beta\hat{H}}$, which is equivalent to the square of the wavefunction at low and to the Maxwell-Boltzmann distribution at high temperature, can be decomposed into a product of density matrices, each at higher effective temperature $\beta_{P}=\beta/P$. As such, Eq.~\ref{PartitionFunction} is a direct consequence of the Trotter theorem and implies that in the limit $P \rightarrow \infty$, sampling $Z_P$ classically is identical to the exact canonical quantum partition function $Z(\beta)$ \cite{parrinello1984moltenkcl}. 


Inserting $P-1$ complete sets of position eigenstates and introducing momenta using the standard Gaussian integral, as well as employing the symmetric Trotter splitting to decompose $\hat{H}$, 
\begin{equation}
  \label{Zp}
  Z_{P}=\mathcal{N} \int d^{NP}\,\mathbf{R}\int
  d^{NP}\,\mathbf{P}\,e^{-\beta H_{P}(\mathbf{R}, \mathbf{P})}
\end{equation}
can be directly sampled by MD. Herein, $\mathbf{P} = \{\mathbf{P}_I\}$ are the momenta of all $N$ particles, while $\mathcal{N}^{-1}=(2 \pi \hbar)^{3NP}$ is an inessential  normalization constant. The so-called ring-polymer Hamiltonian $H_{P}(\mathbf{R}, \mathbf{P})$ that describes the classical system of all $N\times P$ beads of the original $N$ quantum particles, reads as 
\begin{eqnarray}
  \label{Hp}
  H_{P}(\mathbf{R},\mathbf{P}) &=&\sum_{k=1}^{P} \sum_{I=1}^{N} 
    \frac{ \left( \mathbf{P}_{I}^{(k)} \right)^{2}}{2M_{I}} +
    \frac{M_{I}\omega_{P}^{2}}{2} \left ( \mathbf{R}_{I}^{(k)} -
    \mathbf{R}_{I}^{(k+1)} \right)^{2} \nonumber \\ 
    &+&\frac{1}{P} \sum_{k=1}^{P} V( \mathbf{R}^{(k)} ), 
\end{eqnarray}
where $P$ is the number of imaginary-time slices, while $M_i$ are the particle masses and $\omega_{P}=P/\beta=\beta_P^{-1}$ the angular frequency of the harmonic spring potential between adjacent ring-polymer beads. This is to say that the trace of Eq.~\ref{PartitionFunction}, $H_{P}(\mathbf{R},\mathbf{P})$ is isomorphic to a closed classical ring-polymer, thus $\mathbf{R}_{I}^{(P+1)} = \mathbf{R}_{I}^{(1)}$ \cite{chandler1981isomorphism}.  

Eventually, the thermal quantum expectation value 
\begin{equation}
\langle A \rangle = \frac{1}{Z}\text{Tr} \left[ \hat{A} e^{-\beta \hat{H}} \right]
\label{ExpVal}
\end{equation}
of some (usually position-dependent) operator $\hat{A}$ can be directly determined by a PIMD simulation in terms of 
\begin{equation}
\langle A \rangle \simeq \frac{\mathcal{N}}{Z_{P}} \int d^{NP}\mathbf{R} \, d^{NP}\mathbf{P} \,\, A_{P}(\mathbf{R}) e^{-\beta_{P} H_{P}(\mathbf{R}, \mathbf{P})} A_{P}(\mathbf{R}),
\label{KuboExpVal}
\end{equation}
where 
\begin{equation}
A_{P}(\mathbf{R}) = \frac{1}{P} \sum_{k=1}^{P} A(\mathbf{R}^{(k)}).
\label{Ap}
\end{equation}

\subsection{Ring-Polymer Molecular Dynamics}

In the above formulation of PIMD, the trajectories generated by $H_{P}(\mathbf{R},\mathbf{P})$ are a purely fictitious vehicle for sampling the extended phase space and calculating $\langle A \rangle$ via Eq.~\ref{KuboExpVal}; no dynamic information is attached to the corresponding trajectories.
By contrast, the RPMD approach permits to approximately compute even dynamical properties using appropriately chosen real-time correlation functions of the form
\begin{eqnarray}
c_{AB}(t) = \frac{1}{Z} \text{Tr} \left[ e^{-\beta \hat{H}} \hat{A}(0) \hat{B}(t) \right]
\label{Cab}
\end{eqnarray}
within the path-integral formalism \cite{craig:3368, RPMDreview}, where $\hat{A}(0)$ and $\hat{B}(t)$ are Heisenberg-evolved operators, i.e. $\hat{O}(t)=e^{+i \hat{H}t / \hbar} \, \hat{O} \, e^{-i \hat{H}t / \hbar}$. Specifically, 
the PIMD trajectories are used to determine an approximation of the quantum-mechanical Kubo-transformed time-correlation function according to
\begin{eqnarray}
\label{KuboCab}
\tilde{c}_{AB}(t) &=& \frac{1}{\beta Z} \int_{0}^{\beta} d\lambda \, \text{Tr}\left[ e^{-(\beta - \lambda) \hat{H}} \hat{A}(0) e^{-\lambda \hat{H}}  \hat{B}(t) \right] \\
&\approx& \frac{\mathcal{N}}{Z_{P}} \int d^{NP}\mathbf{R} \, d^{NP}\mathbf{P} \,\, e^{-\beta_{P} H_{P}(\mathbf{R}_{0},\mathbf{P}_{0})} A_{P}(0) B_{P}(t), \nonumber
\end{eqnarray}
where the last line defines the RPMD approximation, which is exact in the short-time and harmonic oscillator, as well as high-temperature limit, where Eq.~\ref{KuboCab} reduces to the classical correlation function \cite{craig:3368, habershon2009quantumleakage, RPMDreview}. Accordingly, it is straightforward to see that RPMD complies to classical MD performed in the extended path-integral phase space. The calculated RPMD correlation functions can be subsequently used to compute quantum-dynamical properties; for example, the time integral of the RPMD velocity autocorrelation function $\tilde{c}_{vv}(t)$ allows the determination of quantum diffusion coefficients \cite{miller2005quantumdiffusion, miller2005Hquantumdiffusion}, i.e. 
\begin{equation}
  \label{DiffusionConstant}
  D = \frac{1}{3} \int_0^{\infty}{dt \, \tilde{c}_{vv}(t)}.
\end{equation}

In the related CMD approach \cite{CMD1993}, the center of mass of the ring-polymer (the centroid) moves on the effective potential generated by the thermal fluctuations of the ring-polymer; again, this approach allows the determination of quantum time-correlation functions and the associated quantum-dynamical properties. As the name implies, in CMD only the time-evolution of the of the centroid coordinates 
\begin{equation}
\mathbf{R}_{I}^{(c)} = \frac{1}{P} \sum_{k=1}^{P} \mathbf{R}_I^{(k)}.
\label{Centroid}
\end{equation}
is required, which, on the downside however, entails that the integration time-step has to be rather small. For the purpose to ameliorate the latter and to accelerate CMD simulations, in the partially adiabatic CMD (PA-CMD) scheme \cite{hone:154103}, the effective masses of the ring-polymer beads are chosen so as to recover the correct dynamics of the ring-polymer centroids. Therefore, the elements of the Parrinello-Rahman mass-matrix are selected so that the vibrational modes of all ring-polymer beads, except for the centroid, are shifted to a frequency of 
\begin{equation}
  \label{FreqShift}
  \Omega=\frac{P^{P/(P-1)}}{\beta\hbar}, 
\end{equation}
which allows for integration time-steps close to the ionic resonance limit \cite{comp_quant_eff}.

\subsection{Ring-Polymer Contraction Scheme}

If we assume, as is usually the case, that the evaluation of the interaction potential and the resulting nuclear forces is the most time-consuming part of an atomistic simulation, it is clear from the above that the computational time of a PIMD simulation is around $P$ times greater than is required for the corresponding classical system using conventional MD. 
Relating to this, it has been demonstrated that the computational expense of PIMD simulations employing empirical force-fields can be dramatically reduced using the RPC approach \cite{markland:024105, Markland2008256}. The origin of this scheme is based on the observation that a PIMD simulation of a rigid water model requires around $P=6$ ring-polymer beads to achieve converged quantum mechanical properties \cite{kuharski1085studystructure}, whereas a PIMD simulation of a flexible water model (using similar intermolecular interaction models) needs roughly $P=32$ beads to converge \cite{comp_quant_eff}. This observation immediately suggests that one can develop a scheme where a small number of beads $P^{\prime}$ is used to evaluate the (usually expensive) intermolecular interaction potential $V_{inter}(\mathbf{R})$, while a much larger number of beads $P$ must be used to evaluate the (usually inexpensive) intramolecular potential $V_{intra}(\mathbf{R})$. From this it follows that by exploiting the trivial decomposition $V(\mathbf{R}) = V_{inter}(\mathbf{R}) + V_{intra}(\mathbf{R})$, where $V_{inter}(\mathbf{R})$ can be considered as the slowly varying soft and $V_{intra}(\mathbf{R})$ as the quickly varying hard part of $V(\mathbf{R})$, the total computational cost of the PIMD simulation can be dramatically reduced compared to the standard approach in which the full potential energy function is evaluated on \textit{all} $P$ ring-polymer beads. 

The RPC approach proceeds by first transforming from the coordinate representation to the normal-mode representation of the free ring-polymer 
using 
\begin{equation}
\mathbf{R}_{I}^{(l)} = \frac{1}{P} \sum_{k=1}^{P} C_{kl}^{(P)} \mathbf{R}_{I}^{(k)},
\label{NormalMode}
\end{equation}
where $\mathbf{C}^{(P)}$ is the transformation matrix that diagonalizes the harmonic spring terms in Eq.~\ref{Hp} \cite{tuckerman:5579}. Next, the $P-P^{\prime}$ normal modes with the highest frequencies are discarded; the justification for this being that those modes with high normal-mode frequencies also have the smallest thermal amplitudes, and so, are not strongly coupled to $V_{inter}(\mathbf{R})$, which is typically slowly varying on the length scale of the ring-polymer. Assuming that an odd number of normal modes is retained, such that $P^{\prime} = 2 m^{\prime} + 1$ conforming with the centroid and the $m^{\prime}$ doubly occupied lowest normal modes, one can then transform back into the coordinate representation using the inverse of the normal-mode transformation matrix for the $P^{\prime}$ lowest odes, i.e. 
\begin{equation}
\mathbf{R}_{I}^{(k^{\prime})} = \sum_{l = -m^{\prime}}^{m^{\prime}} C_{k^{\prime}l^{\prime}}^{(P^{\prime})} \mathbf{R}_{I}^{(l)}.
\label{InvNormalMode}
\end{equation}
The overall result of these transformations is that the coordinates of the contracted $P^{\prime}$-bead ring-polymer are given by 
\begin{equation}
\mathbf{R}_{I}^{(k^{\prime})} = \sum_{k=1}^{P} T_{k^{\prime}k} \mathbf{R}_{I}^{(k)},
\label{NormalModeTransformation}
\end{equation}
where the overall transformation matrix is 
\begin{equation}
T_{k^{\prime} k} = \frac{1}{P} \sum_{l = -m^{\prime}}^{m^{\prime}} C_{k^{\prime}l}^{(P^{\prime})} C_{kl}^{(P)}.
\label{TransformationMat}
\end{equation}
Once the computational demanding intermolecular potential $V_{inter}(\mathbf{R})$ has been evaluated at all $P^{\prime}$ beads in the contracted ring-polymer, the potential energy of the full $P$-bead ring-polymer can be recovered as
\begin{equation}
\sum_{k=1}^{P} V(\mathbf{R}^{(k)}) \simeq \frac{P}{P^{\prime}}\sum_{k^{\prime}=1}^{P^{\prime}} V(\mathbf{R}^{(k^{\prime})}), 
\label{RPCpot}
\end{equation}
while the corresponding nuclear forces 
\begin{equation}
\mathbf{F}_{I}^{(k)} \simeq \frac{P}{P^{\prime}} \sum_{k^{\prime}=1}^{P^{\prime}} T_{k^{\prime}k} \mathbf{F}_{I}^{(k^{\prime})},
\label{RPCforce}
\end{equation}
are obtained using the transformation matrix of Eq.~\ref{TransformationMat}, where $\mathbf{F}_{I}^{(k^{\prime})}$ is the force on atom $I$ evaluated on the $k^{\prime}$-th bead of the contracted ring-polymer.

As demonstrated in PIMD simulations of simple empirical water models \cite{Fanourgakis2009, comp_quant_eff, SpuraWater2015, Kessler2015}, the RPC scheme can accurately reproduce the static and dynamic properties of the full $P$-bead PIMD simulation at a fraction of the computational cost, which are only slightly larger than that of analogue classical MD simulations.

\section{Quantum Ring-Polymer Contraction Method}

Unfortunately, in the form outlined above, the RPC scheme cannot be employed directly to \textit{ab-initio} PIMD simulations. The reason for this is that the usage of an \textit{ab-initio} potential $V_{AI}(\mathbf{R})$ does not allow for a straightforward decomposition into intra- and intermolecular potentials, as exploited in empirical force-fields. As a result, it is not clear how one could make use of the RPC scheme in analogy to the above. Nevertheless, it is not generally appreciated that the main idea of the RPC technique is applicable to arbitrary decompositions into hard and soft potentials. The key point of this work is to propose a novel computational approach that we will refer to as q-RPC method, which, in the spirit to the original RPC approach \cite{markland:024105, Markland2008256}, facilitates to perform \textit{ab-initio} PIMD calculations at essentially the same computational cost than conventional \textit{ab-initio} MD simulations. Contrary to the RPC technique, the q-RPC method does not rely on the ability to 
identify intra- and intermolecular contributions, but, as already alluded to above, permits to decompose $V_{AI}(\mathbf{R})$ by the introduction of an appropriately chosen auxiliary potential $V_{aux}(\mathbf{R})$.


To be specific, our approach proceeds as follows. Without approximation, we can add and subtract $V_{aux}(\mathbf{R})$ to the full \textit{ab-initio} ring-polymer Hamiltonian, giving
\begin{eqnarray}
H_{P}^{AI}(\mathbf{R}, \mathbf{P}) &=& \sum_{k=1}^{P} \sum_{I=1}^{N} \frac{ ( \mathbf{P}^{(k)}_{I} )^2}{2 M_{I}} + \frac{M_{I} \omega_{P}^{2}}{2} \left( \mathbf{R}^{(k)}_{I} - \mathbf{R}^{(k+1)}_{I} \right)^{2} \\ 
&+& \frac{1}{P}\sum_{k=1}^{P} \left[ V_{AI}(\mathbf{R}^{(k)}) + V_{aux}(\mathbf{R}^{(k)}) - V_{aux}(\mathbf{R}^{(k)}) \right] \nonumber \\
&=& H_{P}^{aux}(\mathbf{R},\mathbf{P}) + \frac{1}{P} \sum_{k=1}^{P} \left[ V_{AI}(\mathbf{R}^{(k)}) - V_{aux}(\mathbf{R}^{(k)}) \right]. \nonumber
\label{HpAI}
\end{eqnarray}
Here, we have rewritten $H_{P}^{AI}(\mathbf{R}, \mathbf{P})$ by means of an auxiliary $P$-bead ring-polymer Hamiltonian for $V_{aux}(\mathbf{R})$, which is 
\begin{eqnarray}
H_{P}^{aux}(\mathbf{R}, \mathbf{P}) &=& \sum_{k=1}^{P} \sum_{I=1}^{N} \frac{ ( \mathbf{P}^{(k)}_{I} )^2}{2 M_{I}} + \frac{M_{I} \omega_{P}^{2}}{2} ( \mathbf{R}^{(k)}_{I} - \mathbf{R}^{(k+1)}_{I} )^{2} \nonumber \\ 
&+& \frac{1}{P} \sum_{k=1}^{P} V_{aux}(\mathbf{R}^{(k)}), 
\label{HpAux}
\end{eqnarray}
plus a correction term given by the difference between $V_{AI}(\mathbf{R})$ and $V_{aux}(\mathbf{R})$ evaluated on all ring-polymer beads. The key step in our approach is the notion that if the difference potential $\left[ V_{AI}(\mathbf{R}^{(k)}) - V_{aux}(\mathbf{R}^{(k)}) \right]$ varies sufficiently slowly over the length-scale of the ring-polymer, we can use the RPC approximation of Eq.~\ref{RPCpot} to write
\begin{eqnarray}
&&\sum_{k=1}^{P} \left[ V_{AI}(\mathbf{R}^{(k)}) - V_{aux}(\mathbf{R}^{(k)}) \right] \nonumber \\ 
\simeq \frac{P}{P^{\prime}} &&\sum_{k^{\prime}=1}^{P^{\prime}} \left[ V_{AI}(\mathbf{R}^{(k^{\prime})}) - V_{aux}(\mathbf{R}^{(k^{\prime})}) \right], 
\label{qRPCapprox}
\end{eqnarray}
where we again have $P^{\prime} < P$. From this follows that the ring-polymer Hamiltonian for the full $P$-bead ring-polymer can be approximated as
\begin{eqnarray}
\label{HqQrpc}
H_{P}(\mathbf{R}, \mathbf{P}) &\simeq& H_{P}^{aux}(\mathbf{R},\mathbf{P}) \\
&+& \frac{P}{P^{\prime}} \sum_{k^{\prime}=1}^{P^{\prime}} \left[ V_{AI}(\mathbf{R}^{(k^{\prime})}) - V_{aux}(\mathbf{R}^{(k^{\prime})}) \right], \nonumber
\end{eqnarray}
whereas the nuclear forces are given in analogy to Eq.~\ref{RPCforce} as
\begin{equation}
\mathbf{F}_{I}^{AI} = \mathbf{F}_{I}^{aux} + \frac{P}{P^{\prime}} \sum_{k^{\prime}=1}^{P^{\prime}} T_{k^{\prime}k} [\mathbf{F}_{I}^{AI (k^{\prime})} - \mathbf{F}_{I}^{aux (k^{\prime})}].
\label{FiAi}
\end{equation}
In the limit that $V_{aux}(\mathbf{R}) = V_{AI}(\mathbf{R})$, Eqs.~\ref{HqQrpc} and \ref{FiAi} recover the exact full \textit{ab-initio} ring-polymer Hamiltonian. 

In practice, however, the potential energy for the full ring-polymer on potential $V_{AI}(\mathbf{R})$ is approximated as that arising from $V_{aux}(\mathbf{R})$ plus a correction term derived from the difference between $V_{AI}(\mathbf{R})$ and $V_{aux}(\mathbf{r})$ evaluated on a much contracted ring-polymer. Obviously, the most important upshot of the q-RPC method is that the computationally expensive \textit{ab-initio} potential and nuclear forces must only be evaluated on the contracted ring-polymer; this approach, therefore, has the capacity to enable \textit{ab-initio} PIMD simulations based on $V_{AI}(\mathbf{R})$ at a much lower computational cost than a direct \textit{ab-initio} simulation of the full ring-polymer system.

\section{Application to Liquid Water}

To demonstrate the efficiency and assess the accuracy of our novel q-RPC method, we have performed a series of DFT-based \textit{ab-initio} PIMD and RPMD simulations of liquid water and ambient conditions. Due to the fact that we have analyzed the impact of NQE on liquid water in previous works \cite{comp_quant_eff, SpuraWater2015}, we are confine ourselves to validate the present q-RPC method against full \textit{ab-initio} PIMD reference calculations.

\subsection{Computational Details}

To that extent, we considered a cubic supercell consisting of 512 light water molecules subject to periodic boundary conditions. For the q-RPC method, $V_{AI}(\mathbf{R})$ was represented by DFT \cite{RevModPhys.71.1253, RevModPhys.87.897}, while a force-matched potential based on self-consistent charge density-functional based tight-binding (SCC-DFTB) MD simulations was employed to constitute $V_{aux}(\mathbf{R})$ \cite{SpuraWater2015, Karhan2014, PhysRevB.58.7260, Seifert2012}. The latter was selected to mimic the situation that only a relatively inaccurate auxiliary potential is available \cite{DFTBwater2007, DFTBwater2011, Doemer2013}. The eventual \textit{ab-initio} PIMD simulations were performed in the isobaric-isothermal NPT ensemble at ambient conditions using the second-generation Car-Parrinello MD (CPMD) method of K\"uhne et al. \cite{CP2G, kuehnewater2009, CP2Greview}. In this approach, the fictitious Newtonian dynamics of the original CPMD technique \cite{PhysRevLett.55.2471} is replaced with an improved coupled electron-ion dynamics that keeps electrons close to the instantaneous electronic ground state without defining a fictitious mass parameter. The superior efficiency originates from the fact that the iterative wave function optimization is fully bypassed. Thus, not even a single diagonalization step is required, while at the same time allowing for integration time steps up to the ionic resonance limit. The precise settings were identical to those of previous second-generation CPMD studies of water \cite{kuehnewater2009, kuehnewaterinterface2011, kuehne2ptwater2012, kuehnewater2013, kuehnewaterreview2013, Elgabarty2015}. 

All simulations were performed using the mixed Gaussian and plane waves (GPW) code CP2K/Quickstep \cite{GPW, VandeVondele_quickstep}. In the GPW scheme, the Kohn-Sham orbitals are represented by an accurate triple-$\zeta$ Gaussian basis set with one set of polarization functions (TZVP) \cite{vandevondele2007gaussianbasis}, whereas the charge density is expanded in plane waves using a density cutoff of 280~Ry. The interactions between the valence electrons and the ionic cores was described by norm-conserving Goedecker-Teter-Hutter pseudopotentials \cite{GTH1996pseudo, KrackGTH}, while for the exchange and correlation (XC) functional, the Perdew-Tao-Staroverov-Scuseria (TPSS) meta-generalized gradient approximation was employed \cite{tpss}. Van der Waals interactions, which are typically left out by common local and semi-local XC functionals, were taken into account via 
dispersion-corrected atom-centered potentials (DCACPs) \cite{lilienfeld2005dcacp, PhysRevB.75.205131, chun2009importance_vdw, chun2012structure}. Since we are dealing with a disordered system at finite temperature that also exhibits a large band gap, the Brillouin zone is sampled at the $\Gamma$-point only.

The parameters of the q-TIP4P/F-like auxiliary water potential were obtained by matching the nuclear forces to SCC-DFTB reference calculations using the scheme of Spura et al. \cite{SpuraWater2015}. The initial parameters were taken from the original q-TIP4P/F water model \cite{comp_quant_eff}, and reoptimized by minimizing he normalized $L_1$ force distance between the water potential and SCC-DFTB is minimized using the SLSQP algorithm of Kraft \cite{SLSQP}. 

The equations of motion were integrated using a discretized time-step of 0.5~fs. At first, a 2.5~ns long PIMD simulation with $P=32$ ring-polymer beads was conducted in the canonical NVT ensemble using the q-TIP4P/F water model to equilibrate the whole system. Thereafter, a 50~ps long DFT-based \textit{ab-initio} PIMD calculation is performed using q-RPC method with $P = 32 \rightarrow 1$, which means that the full $P=32$ ring-polymer system is computed using $V_{aux}(\mathbf{R})$, while $V_{AI}(\mathbf{R})$ is evaluated on the centroid only, i.e. $P^{\prime} = 1$. From this simulation 25 equally distributed configurations separated by 2~ps each were extracted. All observables were averaged over 25 independent 2~ps long \textit{ab-initio} PIMD calculations using the direct DFT-based PIMD approach, as well as the present q-RPC method. In total, \textit{ab-initio} MD simulations of more than 2~ns have been accumulated. In spite the fact that the second-generation CPMD was employed throughout, which yields a speed-up of roughly a factor of eight \cite{kuehnewater2009, kuehnewaterinterface2011}, to the best of our knowledge, this represents the most extensive \textit{ab-initio} MD simulation conducted to date.

\subsection{Structure of Water at Ambient Conditions}

\begin{figure}
\includegraphics[width=1.0\linewidth]{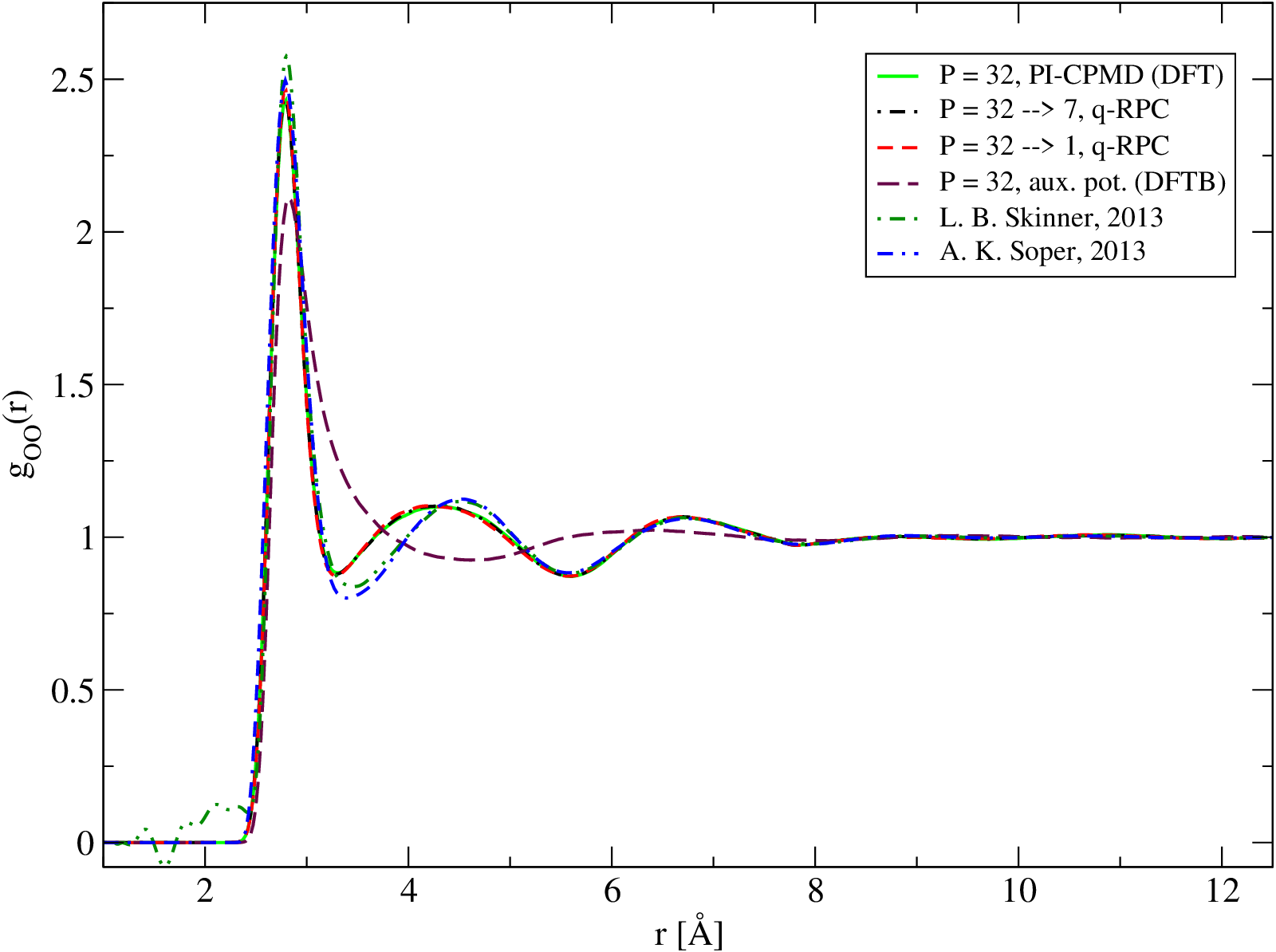}
\caption{Oxygen-Oxygen RDF as obtained by the present q-RPC method and explicit \textit{ab-initio} PIMD reference calculations. The experimental RDFs from Refs.~\onlinecite{soper2013radial} and \onlinecite{skinner2013oo} are shown for comparison.}\label{PimdOO}
\end{figure}

\begin{figure}
\includegraphics[width=1.0\linewidth]{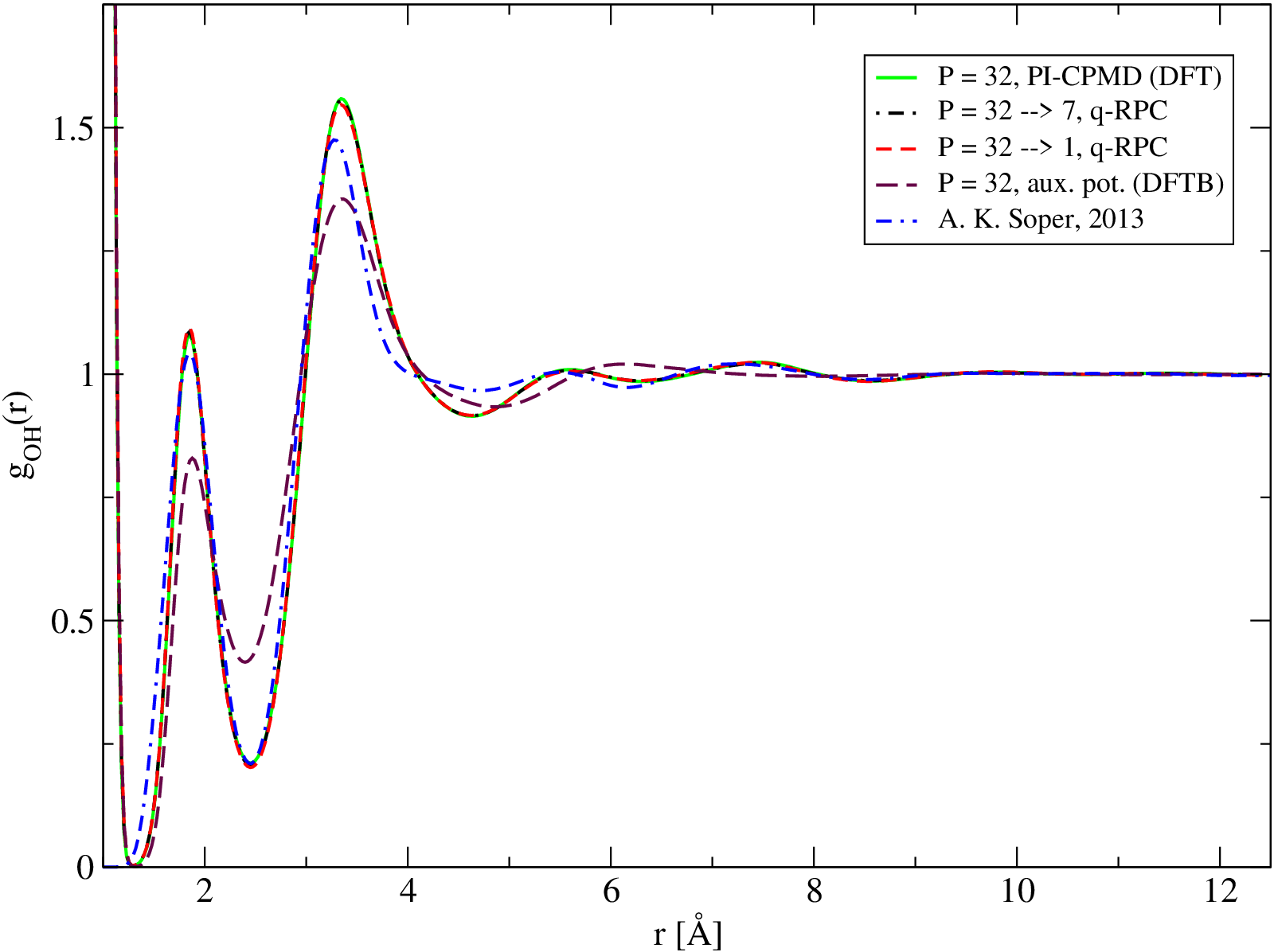}
\caption{Oxygen-Hydrogen RDF as obtained by the present q-RPC method and explicit \textit{ab-initio} PIMD reference calculations. The experimental RDF from Ref.~\onlinecite{soper2013radial} is shown for comparison.}\label{PimdOH}
\end{figure}

\begin{figure}
\includegraphics[width=1.0\linewidth]{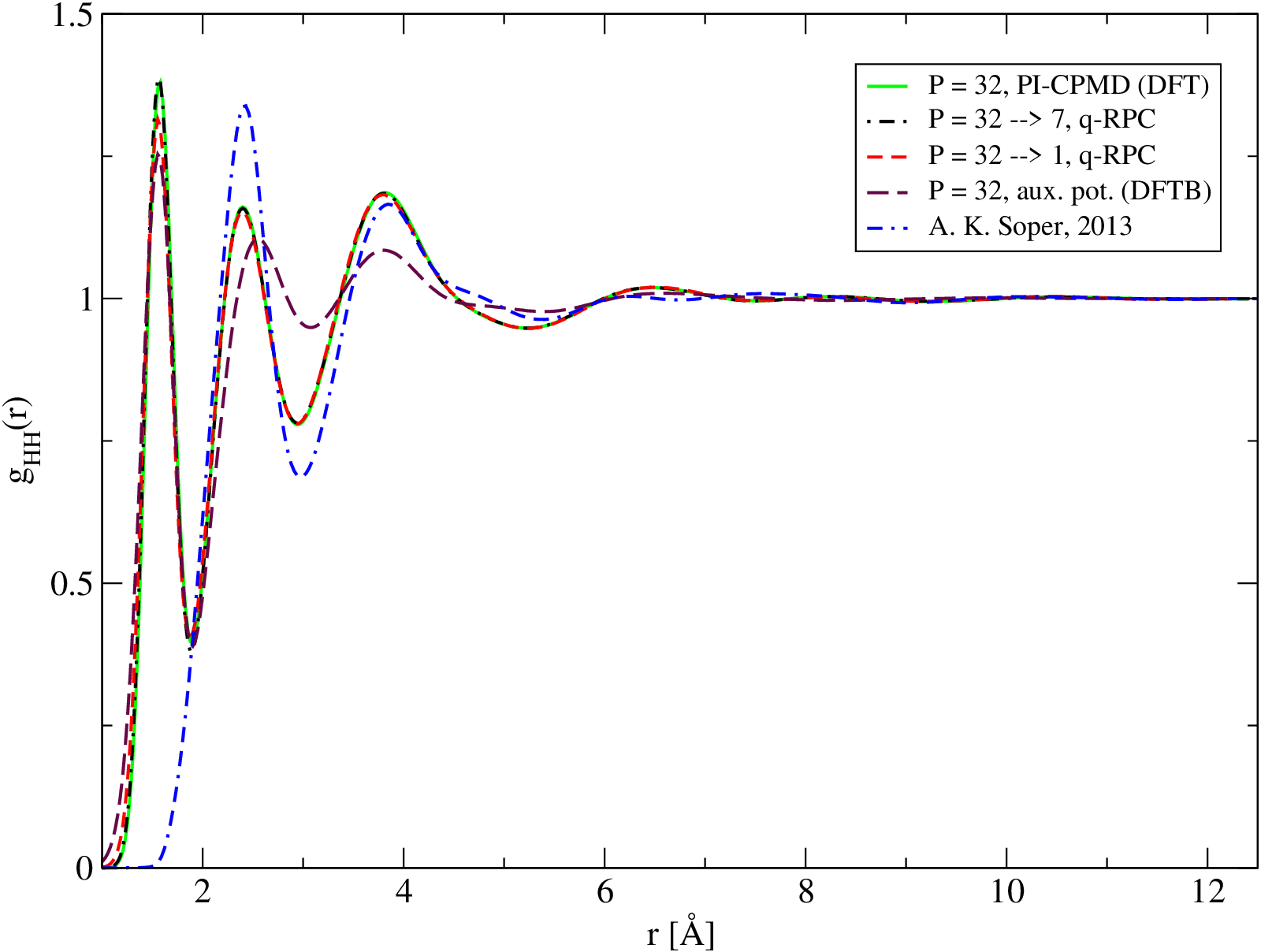}
\caption{Hydrogen-Hydrogen RDF as obtained by the present q-RPC method and explicit \textit{ab-initio} PIMD reference calculations. The experimental RDF from Ref.~\onlinecite{soper2013radial} is shown for comparison.}\label{PimdHH}
\end{figure}

To assess the accuracy of our q-RPC method on the structure of liquid water, the partial radial distribution functions (RDF) are computed \cite{kuehnegofr2013}. The corresponding results are shown in Figs.~\ref{PimdOO} to \ref{PimdHH} and summarized in Table~\ref{TableGofr}, respectively. 
\begin{table}
\caption{Various structural properties of liquid water at ambient conditions as computed using the present q-RPC method and explicit \textit{ab-initio} PIMD reference calculations.}\label{TableGofr}
\begin{ruledtabular}
\begin{tabular}{l c c c c} 
 Method &  $R_{OH}$\,[\AA] 
 &$R_{O\cdot \cdot H}$\,[\AA] 
 &$R_{OO}$\,[\AA]
 \\  \colrule
 32, PI-CPMD (DFT) & 0.978 & 1.854 & 2.796 \\
 $32\rightarrow 7$, q-RPC&0.979&1.851&2.794 \\
 $32\rightarrow 1$, q-RPC&0.980&1.852&2.795 \\
 32, aux. pot. (DFTB) & 0.981 & 1.885 & 2.831 \\
\end{tabular}
\end{ruledtabular}
\end{table}  
It is apparent that the DFTB-based auxiliary potential (aux. pot.) barely resembles the experimental partial RDFs \cite{soper2013radial, skinner2013oo}. Interestingly, however, the partial RDFs of the simulation using the q-RPC method with $P = 32 \rightarrow 1$ are essentially indistinguishable from DFT-based \textit{ab-initio} PIMD reference calculations (PI-CPMD) with $P = 32$ ring-polymer beads. As shown in Table~\ref{TableGofr}, all equilibrium distances are converged within very few thousands of an angstrom, which is an impressive manifestation that the q-RPC method is rather robust with respect to the employed auxiliary potential. In comparison to experimental measurements \cite{soper2013radial, skinner2013oo}, the agreement is generally rather good, except for a somewhat underestimated average distance of the second neighbors in the O-O RDF in Fig.~\ref{PimdOO}, which corresponds to the first solvation shell of liquid water. Moreover, the absolute height of the first intermolecular peak of the H-H RDF in Fig.~\ref{PimdHH} is slightly too low.

\subsection{Kinetic and Potential Energy}

A particular stringent test to assess the convergence of PIMD simulations are the average potential and in particular the kinetic energy per water molecule. The latter can be computed from the centroid virial estimator \cite{HermanJCP1982, CeperleyRMP}
\begin{equation}
 \left< E_{kin} \right> = \frac{3}{2NP} \sum_{I=1}^{N} { \sum_{k=1}^{P} { (\mathbf{R}_{I}^{(k)} - \mathbf{R}_{I}^{(c)}) \frac{\partial V(\mathbf{R}^{(k)})}{\partial \mathbf{R}_{I}^{(k)}}}} + \frac{9}{2\beta}, 
\end{equation}
where $\mathbf{R}_{I}^{(c)}$ is computed according to Eq.~\ref{Centroid}, while $9/(2\beta)$ is the classical kinetic energy.

\begin{table}
\caption{The average potential and kinetic energy per water molecule, as well as the translational diffusion constant, as determined by the present q-RPC method and explicit \textit{ab-initio} PIMD reference calculations.}\label{TableE}
\begin{ruledtabular}
\begin{tabular}{l c c c} 
 Method &  $\langle V_{mol} \rangle$\,[$\frac{kJ}{mol}$] &$\langle
 E_{kin}\rangle$\,[$\frac{kJ}{mol}$]& D\,[$\frac{\text{\AA}^2}{ps}$] \\ \colrule
 32, PI-CPMD (DFT) &-8.10&32.10&0.432 \\ 
 $32\rightarrow 7$, q-RPC &-8.56&31.79&0.421 \\
$32\rightarrow 1$, q-RPC&-9.49&30.88&0.403 \\
32, aux. pot. (DFTB) & -31.05&11.14&1.037 \\
\end{tabular}
\end{ruledtabular}  
\end{table} 
The average potential energy $\left< E_{pot} \right>$ and $\left< E_{kin} \right>$ as obtained by the present PIMD simulations can be found in Table~\ref{TableE}. Despite the rather approximate nature of $V_{aux}(\mathbf{R})$, using the q-RPC method and with $P = 32\rightarrow 1$, both $\left< E_{pot} \right>$ and $\left< E_{kin} \right>$ are converged within less than 1.4~kJ/mol, while for $P = 32\rightarrow 7$ the error is smaller than 0.5~kJ/mol. As a consequence, the former is probably good enough for most purposes, whereas the latter can be considered as fully converged.

\subsection{Translational Diffusion Coefficient}

\begin{figure}
\includegraphics[width=1.0\linewidth]{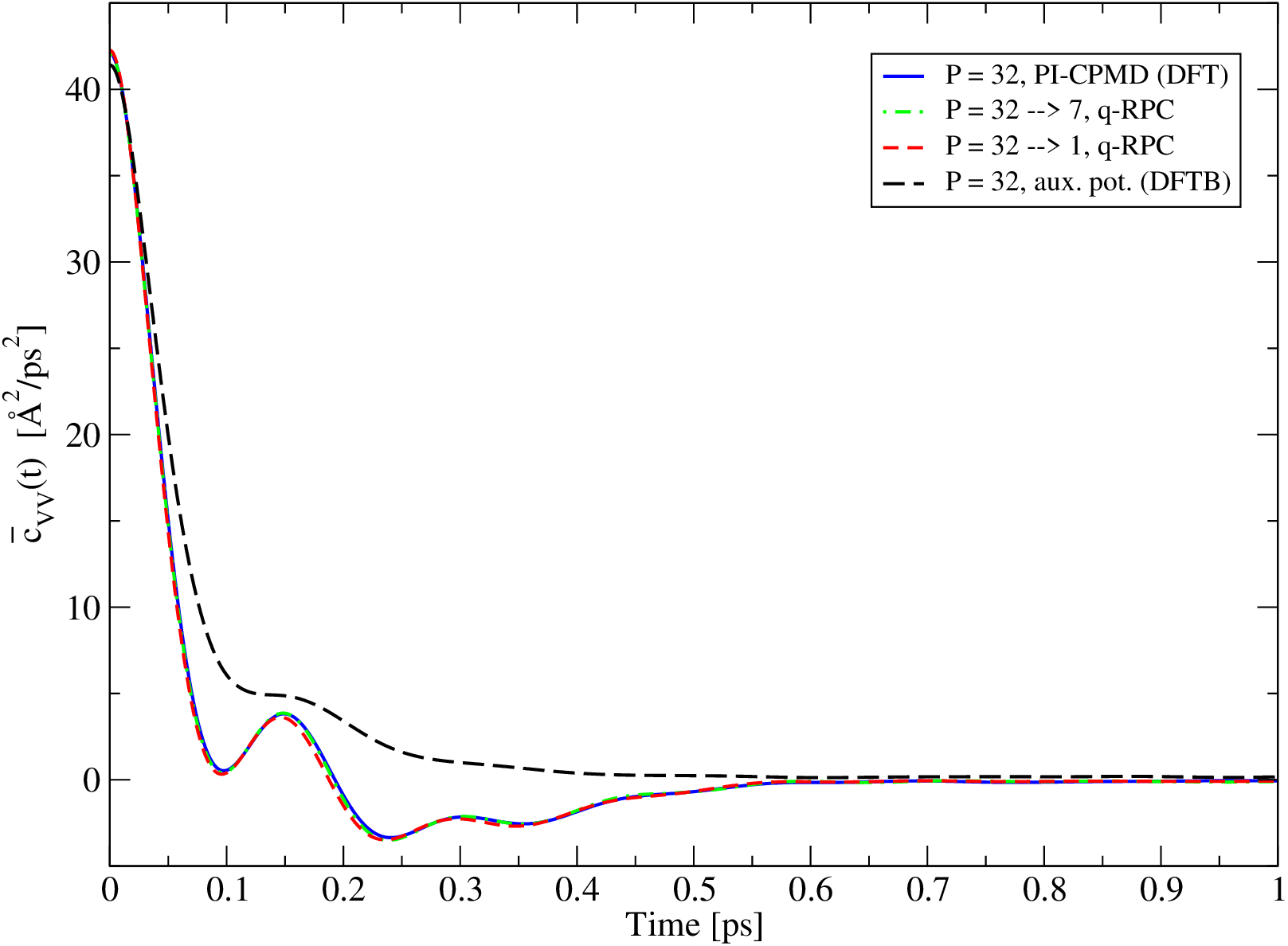}
\caption{Velocity autocorrelation function as obtained by the \textit{ab-initio} RPMD simulations using the present q-RPC method and explicit \textit{ab-initio} PIMD simulations.}\label{Cvv}
\end{figure}

The RPMD velocity autocorrelation functions $\tilde{c}_{vv}(t)$ are displayed in Fig.~\ref{Cvv}. As can be seen, applying the q-RPC method, $P = 32\rightarrow 1$ is sufficient the obtain essentially indistinguishable results in comparison to DFT-based \textit{ab-initio} PIMD reference calculations with $P = 32$ ring-polymer beads. From this it follows that even dynamical properties are equally rapidly converging than structural properties and already $P = 32\rightarrow 1$ practically fully converged. 

The associated translational quantum diffusion coefficients $D$ were calculated via Eq.~\ref{DiffusionConstant} and are printed in Table~\ref{TableE}. Strikingly, the diffusion constant of the auxiliary potential is more than a factor 4 higher than the experimental value of 0.23 \AA$^2$/ps \cite{Price1999}, but in excellent agreement with previous studies using SCC-DFTB \cite{Maupin2010, Karplus2011, DFTBwater2011}. Nevertheless, this well-know deficiency of the SCC-DFTB approach has been mitigated by the recently proposed DFTB3 scheme \cite{DFTBwater2011, Gaus2012}. The present \textit{ab-initio} PIMD simulations also overestimate the experimental value for $D$ even though to a much smaller extent.

\section{Conclusion}

In summary, we have presented a novel computational approach, dubbed q-RPC method, which enables to decompose an arbitrary \textit{ab-initio} potential into a rapidly varying hard and a more slowly varying soft part via a computationally inexpensive auxiliary potential. To further reduce the computational expense, the latter can be computed using the original RPC scheme, resulting in, what we call, double RPC (d-RPC) method. Using the example of liquid water, it was demonstrated that the q-RPC method permits for \textit{ab-initio} PIMD simulations at essentially no extra computational cost with respect to conventional \textit{ab-initio} MD calculations. 

We conclude by noting that the present q-RPC method is identical to a multiple time step algorithm in imaginary-time \cite{TuckermanMTS1991, RESPA}. In fact, an alternative derivation based on the Liouville operator formalism immediately suggest to devise a multiple time step scheme in imaginary- \textit{and} real-time \cite{CeriottiMTS2015, MarklandMTS2015}. A corresponding approach using coupled-cluster and Hartree-Fock methods for $V_{AI}(\mathbf{R})$, and $V_{aux}(\mathbf{R})$, respectively, will be presented elsewhere.

\begin{acknowledgments}
Financial support from the Graduate School of Excellence MAINZ and the IDEE project of the Carl Zeiss Foundation is kindly acknowledged. Moreover, we would like thank the Gauss Center for Supercomputing (GCS) for providing computing time through the John von Neumann Institute for Computing (NIC) on the GCS share of the supercomputer JUQUEEN at the J\"ulich Supercomputing Centre (JSC). Parts of this research were conducted using the supercomputer OCuLUS of the Paderborn Center for Parallel Computing (PC$^2$), as well as the supercomputer Mogon of the Johannes Gutenberg University Mainz, which is a member of the AHRP and the Gauss Alliance e.V.
\end{acknowledgments}

\bibliography{ts} 

\end{document}